\documentclass{myws-mpla}
\usepackage{wrapfig}     

\newcommand{\W}{\mbox{$W$}}

\newcommand{\GeV}{\mbox{\rm ~GeV}~}
\newcommand{\GeVx}{\mbox{\rm ~GeV}}

\newcommand{\GeVsq}{\mbox{${\rm ~GeV}^2$}~}
\newcommand{\GeVsqx}{\mbox{${\rm ~GeV}^2$}}

\newcommand{\pb}{\mbox{${\rm ~pb}$}~}
\newcommand{\pbx}{\mbox{${\rm ~pb}$}}

\newcommand{\fbx}{\mbox{${\rm ~fb}$}}

\newcommand{\pbinv}{\mbox{${\rm ~pb^{-1}}$}}

\newcommand{\lsim}{\raisebox{-0.5mm}{$\stackrel{<}{\scriptstyle{\sim}}$}}

\newcommand{\rp}{$\not$\kern-0.pt$R_p$}

\def\z0{Z^0}

\def\ept{ep\rightarrow e t X}
\def\kg{\kappa_{tu\gamma}}
\def\mt{M_{\rm top}}
\def\vz{v_{tuZ}}

\begin{document}
\markboth{T. Carli, D. Dannheim, L. Bellagamba } {Isolated Leptons and Large Missing Transverse 
Momentum at HERA}
%
%
\catchline{}{}{}{}{}
%
%
\title{Events with Isolated Charged Leptons and Large Missing Transverse 
Momentum at HERA}
\author{\footnotesize Tancredi Carli}
\address{CERN, Experimental Physics Division, CH-1211 Geneva 23, Switzerland \\
Tancredi.Carli@cern.ch}
\author{Dominik Dannheim}
\address{DESY, Notkestr. 85, D-22607 Hamburg \\
Dominik.Dannheim@desy.de}
\author{Lorenzo Bellagamba}
\address{INFN, Dipartimento di Fisica, via Irnerio 46, I-40126 Bologna  \\
Lorenzo.Bellagamba@bo.infn.it}
\maketitle
%
%
\maketitle

\begin{abstract}
Striking events with isolated charged leptons, large missing transverse
momentum and large transverse momentum of the hadronic final state
($P_T^X$) were observed at the electron proton collider
HERA in a data sample
corresponding to an integrated
luminosity of about $130$\pbinv. The H1 collaboration observed $11$ events with
isolated electrons or muons and with $P_T^X >25 \GeVx$.
Only $3.4 \pm 0.6$ events were expected from
Standard Model (SM) processes. Six of these
events have $P_T^X >40 \GeVx$, while $1.3 \pm 0.3$ events were
expected. The ZEUS collaboration observed good agreement with the SM.
However, ZEUS found two events with a similar
event topology, but tau leptons instead of electrons or muons in the final state. 
Only $0.2 \pm 0.05$ events were expected from  SM processes.
For various hypotheses the compatibility of the experimental results
was investigated with respect to the SM and with respect to
possible explanations
beyond the SM. Prospects for the high-luminosity HERA-II data taking period
are given.
\end{abstract}

%
%
\section{Introduction}
The Standard
$SU(3) \otimes SU(2) \otimes U(1)$
Model (SM) of particle physics describes the electroweak and
strong interactions between elementary particles in both the
low- and the high-energy regime to an amazing accuracy.
It is, however, unsatisfactory in the sense that many
fundamental facts such as the quark-lepton symmetry,
the structure of the gauge groups and the spectrum of the particle masses
remain unexplained. Furthermore, the inclusion
of gravitation as the fourth fundamental force in nature
remains an open question.
Recently, hints for the need of an extension of the SM
were obtained by the observation of neutrino
oscillations suggesting a flavour mixing also in the
leptonic sector and non-zero neutrino rest masses\cite{neu_oszis}. 
Only future 
experiments will be able to fully clarify
the nature of this phenomenon.
Anyway, a crucial step forward, towards a deeper understanding of
the fundamental particles and their interactions, would be the
experimental observation of new heavy particles beyond the
presently known particle spectrum.

The high centre-of-mass energy, $\sqrt{s}$,
of HERA colliding positrons of energy $E_e = 27.5$~GeV
with protons of energy $E_p = 920$~GeV
offers the possibility to directly produce 
new heavy particles with a mass up to 
$\sqrt{s}=318$~GeV\footnote{In the data taking period
$1994$-$1997$ the proton energy was $820$\GeV ($\sqrt{s}=300$~\GeVx).
This period corresponds to an integrated luminosity of 
$37$\pbinv ($48$\pbinv) for H1 (ZEUS). In
$1998/1999$ HERA collided electrons on protons, corresponding to an
integrated luminosity of approximately $14$\pbinv~ ($17$\pbinv) for H1 (ZEUS).}.
In addition, new heavy particles can also interfer, via virtual effects,
with SM processes, resulting in an enhancement or deficit 
in measured cross sections over the SM expectation.
HERA is hence also sensitive to particles with
masses higher than the centre-of-mass energy. 

Searches for new particles and new interactions beyond the SM (BSM) 
at HERA were reviewed recently\cite{siroiskuze}. 
Here, the final HERA-I results\cite{h1,h1-singletop,ZEUS00,ZEUS03,ZEUStau}  
on the search for events 
with isolated charged leptons ($l$: electron, muon or tau lepton),
with large missing transverse momentum ($P_T^{\rm miss}$) and
with large transverse momentum of the hadronic final state ($P_T^X$),
$ep \to l\nu X$, are reviewed. 
Such an event signature is typical for a singly produced heavy particle
decaying into a charged lepton and a neutrino. 
The possibility to produce BSM particles leading to this signature
has been extensively discussed\cite{kon,pl:b457:186,Rodejohan}.

\section{Calculation of the SM Process  $ep \to e W^\pm X$}
\begin{wrapfigure}[17]{l}{7.5cm}
\vspace{-0.5cm}
\psfig{file=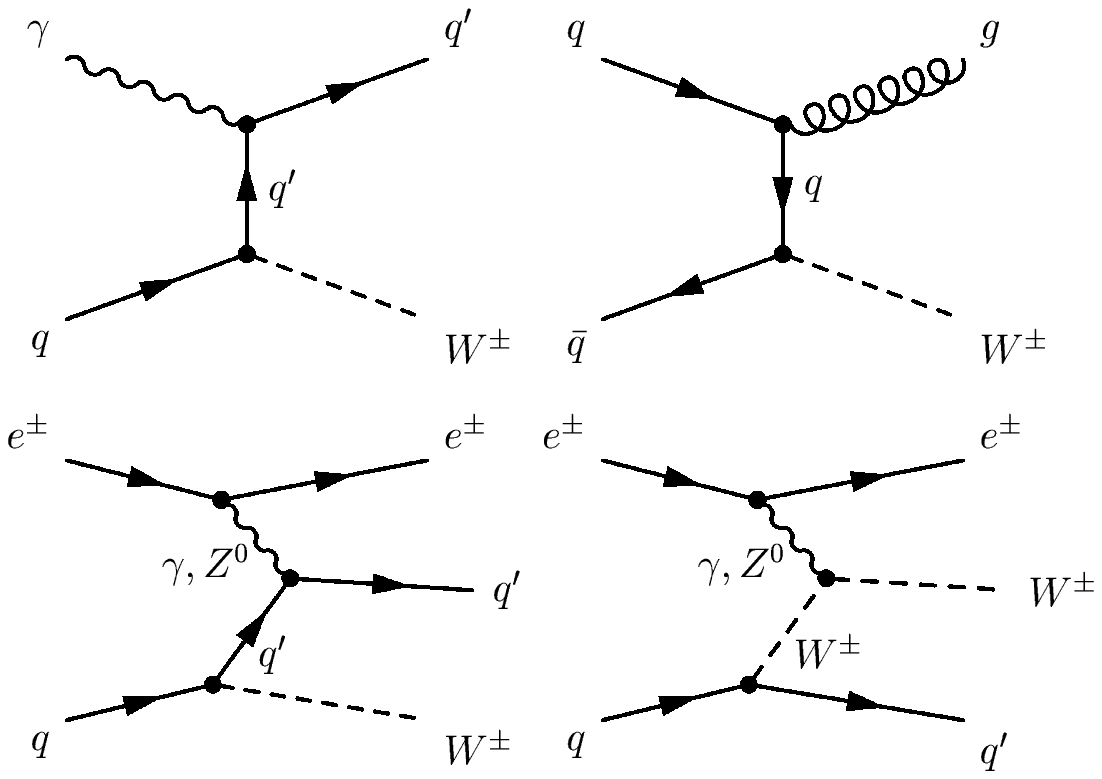,width=7.5cm}
\begin{picture}(0,0) 
\put(5,130){a)} \put(115,130){b)}  
\put(5,45){c)} \put(115,45){d)}
\end{picture}
\vspace*{-8pt}
\caption{Examples of leading order Feynman diagrams 
for $W^\pm$-boson production at HERA through direct (a) and resolved (b)
photoproduction, the DIS process (c) and triple gauge boson coupling (d).
\label{fig:w_prod}}
\vfill
\end{wrapfigure}
Within the SM the dominant process leading to 
isolated charged leptons and large $P_T^{\rm miss}$
is the production of $W^\pm$-bosons ($ep \to e W^\pm X$).
This process has a small cross section. 
The inclusive $W^\pm$-boson production cross section is 
$\sigma(ep \to e W^\pm X) \approx 1 \pbx$. However, in this process 
the transverse momentum of the hadronic final state $P_T^X$
is expected to be relatively small. If $P_T^X > 25 \GeV$ is required,
the cross section drops to approximately $0.2$\pbx.

Typical
leading order Feynman diagrams of $W^\pm$-boson production at HERA
are shown in Fig.~\ref{fig:w_prod}.
The first two diagrams represent collisions of real photons with
protons (photoproduction). In Fig.~\ref{fig:w_prod}a the photon
interacts directly as point-like particle $\gamma q \to q W^\pm$ (direct photoproduction), 
in Fig.~\ref{fig:w_prod}b it splits into a quark anti-quark pair
before interacting with the hard subprocess $q q \to g W^\pm$ (resolved photoproduction). 
Figure~\ref{fig:w_prod}c shows the deep-inelastic scattering (DIS) process $\gamma^* q \to q W^\pm$, 
where the photon is virtual. If the photon virtuality is high 
($Q^2 \gg 1 \GeVsqx$), 
the scattered electron can be measured in the main detector. 
In each of these diagrams, the $W^\pm$-boson is radiated from a quark
in the proton.
This is the dominant single contribution of in total seven contributing
diagrams.
Of particular interest is
the diagram where the interaction proceeds via the triple gauge boson
coupling ($\gamma W^\pm W^\pm$, see Fig.~\ref{fig:w_prod}d). This process 
allows the
anomalous trilinear coupling of gauge bosons to be tested at HERA\cite{baur89}.
However, limits on an anomalous triple boson coupling obtained
on part of the data set by ZEUS\cite{ZEUS00} are not competitive with the 
precise data from LEP\cite{triplelep} and Tevatron\cite{tripletevatron}.

The direct photoproduction process gives the dominant contribution
to the total cross section.
The DIS contribution is about a factor of two smaller.
The resolved photoproduction process 
is negligible at high transverse momenta. The process $ep \to  \nu W^\pm X$
contributes about $5\%$ and can thus be neglected.

The leading order (LO) ${\cal O}(\alpha^2)$ 
calculation and Monte Carlo (MC) simulation of the $ep \to e W^\pm X$ process
has been available since the start-up of HERA\cite{epvec,diaconu}.
Recently, the ${\cal O}(\alpha^2 \alpha_s)$ QCD corrections
were calculated for the dominant direct photoproduction contribution\cite{dss}.
They include the virtual corrections due to 
one-loop diagrams generated by virtual gluon
exchange and the real corrections due to gluon radiation from the
quark lines. In the calculation, the renormalisation and factorisation
scales were set to $\mu^2 = M_W^2$, where $M_W$ is the mass
of the $W^\pm$-boson. The QCD corrections modify the LO
result by $ (10-15\%)$ and reduce the dependence of the
calculated cross section on  $\mu$ from about $20\%$ to
about  $5 \%$\footnote{The residual scale dependence is evaluated
by varying $\mu$ in the range $4 \mu^2 - 0.25 \mu^2$.}.
For the analyses reviewed here, a reweighting method for the LO
$W^\pm$-production MC simulations was used\cite{w-reweighting}, 
which takes into account the QCD corrections and reduces the
uncertainty on the cross section to approximately 10-15\%.

\section{Observation of Events with Isolated Electrons or Muons}
\label{sec:isolated_emu}
Events were selected by requiring 
large $P_T^{\rm miss}$ 
and an electron (e) or muon ($\mu$) 
with high transverse momentum ($P_T^l$)
in the acceptance of the 
detector.
The lepton ($l$) had to be isolated, i.e. the distance in the $\eta-\phi$ 
plane\footnote{The variable $\eta$ denotes the pseudo-rapidity
defined by $\eta=-\ln{\tan{(\theta/2)}}$, where $\theta$ is the polar
angle measured with respect to the proton-beam direction. The 
variable $\phi$ denotes the azimuthal angle.}
from the axis of the closest jet
$D_{\rm jet} = \sqrt{\Delta \eta_{l{\rm jet}}^2 + \Delta \phi_{l{\rm jet}}^2}$
and the distance from the nearest track  
$D_{\rm trk} = \sqrt{\Delta \eta_{l{\rm trk}}^2 + \Delta \phi_{l{\rm trk}}^2}$ had to be large. 
Jets were defined by the inclusive longitudinally invariant $k_T$ 
clustering algorithm\cite{inclkt} and were required to have  
a transverse energy $E_T > 5 \GeVx$. ZEUS required in addition $-1 <\eta_{\rm jet}<2.5$.
In the electron channel, H1 applied the cut on $D_{\rm trk}$ only for electron
polar angles bigger than $45^\circ$.

To efficiently remove neutral current (NC) DIS, events with a back-to-back
topology in the azimuthal plane were rejected by a cut on the
difference between the direction of the lepton and the hadronic final state momentum
$\Delta \Phi_{lX}$. Different cut values were applied for the electron and muon 
channel. 
Furthermore a cut on 
the longitudinal momentum balance $\delta_{\rm miss}$ was applied\footnote{
The variable $\delta_{\rm miss}$ is defined as 
$\delta_{\rm miss}= 2 E_e - \sum_i E_i (1 - \cos{\theta_i})$, where
$E_i$ and $\theta_i$ denote the energy and the polar angle of each 
energy deposit, respectively, 
and $E_e$ is the electron beam energy. 
For an event where only momentum in the proton-beam direction is undetected, 
$\delta_{\rm miss} =0$.}.

To gain sensitivity at low $P_T^{\rm miss}$, 
H1 used the ratio $V_{\rm ap}/V_{\rm p}$ 
of the anti-parallel to parallel components of the
measured calorimetric transverse momentum, with respect to
the direction of the calorimetric transverse momentum.
This variable measures the azimuthal balance of the event. 
Events with high-$p_T$ particles that do not deposit much energy in
the calorimeter ($\mu$, $\nu$) generally have low values of 
$V_{\rm ap}/V_{\rm p}$.
In addition, in the electron channel
the reconstructed squared momentum transfer
$\xi_e^2=4 E'_e E_e \cos{\theta_e}/2 > 5000 \GeVsqx$ was used
for $P_T^X < 25 \GeVx$, where
$E'_e$ is the energy and $\theta_e$ the polar angle of the
isolated electron. This observable corresponds to the photon 
virtuality $Q^2$, if the scattered electron in a
NC DIS process is measured. Since the NC DIS cross section
falls steeply with $Q^2$, $\xi_e^2$ is generally lower in NC DIS than in 
$W^\pm$-boson production.
%

\begin{wraptable}[13]{l}{7.5cm}
\vspace{-0.4cm}
\tbl{\label{tab:iso_cut}
Main H1 and ZEUS event selection cuts for events
with isolated electrons ($e$) or muons ($\mu$).
Some of the cuts are only applied in the $e$
or $\mu$ channel. 
}{
\begin{tabular}{c|c}
H1    & ZEUS \\
\hline
$5^{\circ} < \theta_l < 140^\circ$   & $17^\circ \lsim \, \theta_l < 115^\circ$ \\
$P_T^l > 10 \GeV$    & $P_T^l > 5 \GeV$ \\
$P_T^{\rm miss} > 12 \GeV$ & $P_T^{\rm miss} > 20 \GeV$ \\
$D_{\rm jet} > 1.0$ & $D_{\rm jet} > 1.0$ \\ 
$D_{\rm trk} > 0.5$ for $\theta_e > 45^\circ$ (e)& $D_{\rm trk} > 0.5$ \\ 
$D_{\rm trk} > 0.5$ ($\mu$)&  \\ 
$\Delta \phi_{lX} < 160^\circ ($e$), 170^\circ (\mu)$ & $\Delta \phi_{lX}<172^\circ$ ($e$)\\
$\delta_{\rm miss} > 5 \GeV$ & $\delta_{\rm miss} > 8 \GeV$ ($e$) \\ 
$V_{ap}/V_{p}<0.5$ ($e$) & \\
$\xi_e^2 > 5000 \GeVsq$ ($e$) & \\
\end{tabular}
}
\end{wraptable}
The most important selection criteria for the H1 and ZEUS
analyses are summarised in Tab.~\ref{tab:iso_cut}. The main differences
are the requirements on  $P_T^{\rm miss}$ and the lepton acceptance. 

These selection criteria were designed to reject the main
background processes with high cross sections, i.e.
mismeasured NC, $ep \to e X$, 
and charged current (CC), $ep \to \nu X$,  
DIS events, two-jet-photo\-pro\-duction events, $ep \to {\rm jet} \, {\rm jet} \, X$,
and events where two leptons are produced in inelastic
photon-photon collisions,
$ep \to e l^+ l^- X$.
The H1 event selection has been optimised to increase the acceptance
for $W^\pm$-boson
production, resulting in a selection efficiency of $40\%$ for
$ep \to W^\pm X$ events with $P_T^X > 25 \GeVx$.
The ZEUS event selection was more oriented towards
the search for a singly produced heavy particle and was less efficient
for $W^\pm$-boson production.

\begin{table}[th]
\tbl{Number of measured and expected events in the 
H1 and ZEUS analyses of events with isolated electrons or muons,
large $P_T^{\rm miss}$ and large $P_T^X$. 
Both, the number of 
events expected from all SM processes and the contribution from the $e p \to e W^\pm X$ process in percent are given.
\label{tab:iso_h1_results_el_mu}}
{
\begin{tabular}{c|ccc|ccc}
  \multicolumn{1}{c}{}& \multicolumn{3}{c}{$P_T^X > 25 \GeV$} &  \multicolumn{3}{c}{$P_T^X > 40 \GeV$}  \\
   H1  & Data & SM  & $W^\pm$-contr. & Data & SM  & $W^\pm$-contr. \\
\hline
electron & $5$  &  $1.8 \pm 0.3$ &  $82\%$ & $3$ & $0.7 \pm 0.1$ & $80\%$ \\
muon     & $6$  &  $1.7 \pm 0.3$ &  $88\%$ & $3$ & $0.6 \pm 0.1$ & $92\%$ \\
combined & $11$ &  $3.4 \pm 0.6$ &  $85\%$ & $6$ & $1.3 \pm 0.3$ & $86\%$ \\
\hline
 ZEUS    & Data & SM  & W-contr. & Data & SM  & W-contr. \\
\hline
electron & $2$  &  $2.9 ^{+0.6}_{-0.3}$ &  $45\%$ & $0$ & $0.9 \pm 0.1$ & $61\%$ \\
muon     & $5$  &  $2.8 \pm 0.2$ &  $50\%$ & $0$ & $1.0 \pm 0.1$ & $61\%$ \\
combined & $7$  &  $5.7 \pm 0.6$ &  $47\%$ & $0$ & $1.9 \pm 0.2$ & $61\%$ \\
\end{tabular}
}
\end{table}

\begin{figure}[th]
\begin{center}
\psfig{figure=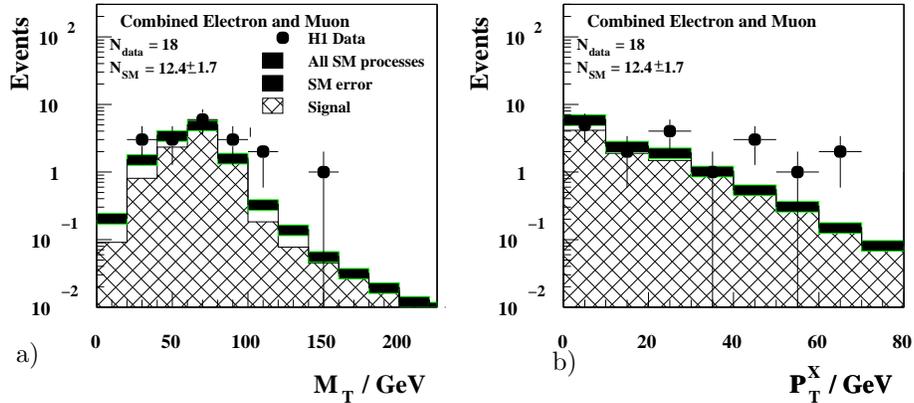,width=12.cm}
\end{center}
\begin{picture}(0,0) 
\put(12, 35){a)} \put(215, 32){b)}  
\end{picture}
\vspace{-1.cm}
\caption{
The transverse mass (a) and transverse momentum of the
hadronic final state (b) distribution for the H1
events with isolated electrons or muons and missing transverse momentum
in the $e^+ p$ data sample ($\int {\cal L} dt \approx$ 105 \pbinv).
The open histogram indicates the expectation for SM processes,
the shaded band the total uncertainty. The hatched
histogram indicates the contribution from  $W^\pm$-boson production. 
The total number of observed
data events ($N_{\rm data}$) and the total number of expected
SM events ($N_{\rm SM}$) is also given.
\label{fig:h1_iso_el_mu}}
\end{figure}

\begin{figure}[th!]
\begin{center}
\psfig{file=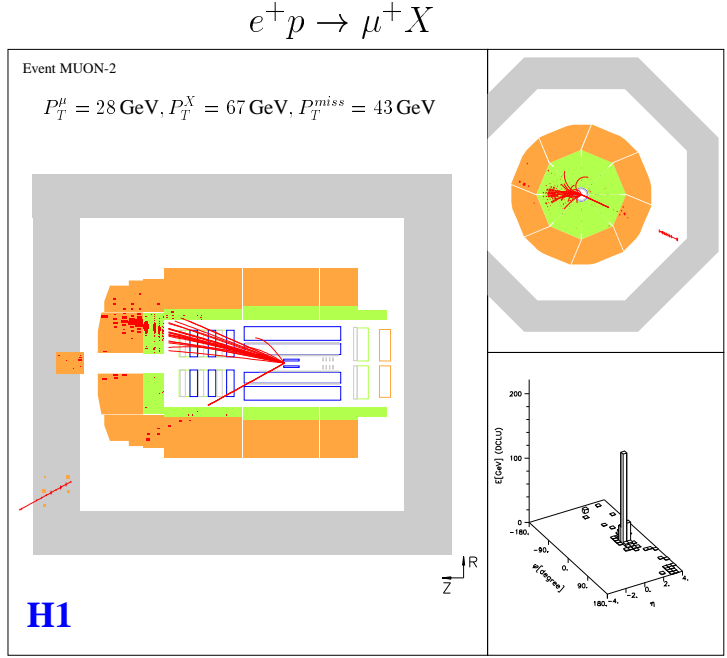,bbllx=62,bblly=225,bburx=581,bbury=702,clip,width=10.cm}
\caption{Display of an isolated muon event with
large $P_T^{\rm miss}$ and large $P_T^X$ in the H1 detector.
\label{fig:h1_muon_event}}
\end{center}
\end{figure}

H1 observed $11$ events in the
electron channel and $8$ events in the muon channel. Most of these
events where found in the $e^+ p$ data taking period corresponding
to $\int {\cal L} dt \approx$ 105 \pbinv.
In the $e^- p$ data, which are a fraction of about $10\%$ of the total
data sample, only one electron event was found.
In the electron (muon)
channel $11.5 \pm 1.5$ ($2.9 \pm 0.5$) events were expected from 
SM processes, $71\%$ ($86\%$) from $e^\pm p \to W^\pm X$.

The transverse-mass ($M_T$) distribution of the electron or muon
and the neutrino reconstructed from the missing momentum
(see Fig.~\ref{fig:h1_iso_el_mu}a) 
is compatible with the distribution expected from a $\W^\pm$-decay
and agrees with the SM expectation.
The striking feature of these events, however, is their large $P_T^X$. 
While for low $P_T^X$ the number
of measured events roughly corresponded to the number of expected events,
an excess of data events was seen towards large $P_T^X$ (see Fig.~\ref{fig:h1_iso_el_mu}b).
The number of measured and expected events with electrons or muons
and with $P_T^X > 25 \GeV$ or $P_T^X > 40 \GeV$ are summarised in
Tab.~\ref{tab:iso_h1_results_el_mu}.
An additional prominent event\cite{H1_the_event,h1_98,H1susy},
where a scattered muon balances
the transverse momentum of the hadronic system, was rejected
by the cut on $\Delta \phi_{lX}$. 

In the electron channel with $P_T^X>25 \GeVx$,
four events had a clearly identified positron ($e^+$), in one event
the charge could not be measured.
In the muon channel three events had a
$\mu^+$ and two a $\mu^-$. For the remaining event the charge could
not be determined with sufficient accuracy.
One of the muon events is shown in Fig.~\ref{fig:h1_muon_event}.

In the combined electron and muon channel, H1 observed $6$ events
with $P_T^X > 40 \GeVx$, while only $1.3 \pm 0.3$ events were expected
from SM processes.
The Poisson probability for the SM expectation to fluctuate
to this observed number of events or more is about $0.3\%$. 

The ZEUS data were 
consistent with the expected SM background in all kinematic regions. 
In the electron (muon) channel 24 (12) events were observed,
while $20.6^{+1.7}_{-4.6}$ ($11.9^{+0.6}_{-0.7}$) events were
expected from SM processes.
For $P_T^X > 25 \GeVx$, $7$ events were found and $5.7 \pm 0.6$ were expected from SM processes. 
No event with $P_T^X > 40 \GeV$ was observed, while $1.9 \pm 0.2$ events were
expected. The excess reported by H1 was thus not confirmed by ZEUS.

\section{Observation of Events with Isolated Tau Leptons}
\label{sec:isolated_tau}
The ZEUS collaboration also extended the search to
events with isolated tau leptons, large $P_T^{\rm miss}$
and large $P_T^X$. 
\subsection{Tau-Lepton Identification}
The tau-leptons were identified in their hadronic decay mode. 
In contrast to jets initiated by
quarks or gluons, jets produced by tau-leptons are
pencil-like, collimated and have a low charged-particle multiplicity.
The tau-lepton identification was developed using independent event 
samples\footnote{The event selection was mainly
based on the requirement of large $P_T^{\rm miss}$.
Details can be found elsewhere\cite{dominik}.}.

The main background process is CC DIS where a jet from the hadronic final state
is misidentified as tau lepton.
The CC DIS cross section is about three orders of magnitude higher
than the signal from  $W^\pm$-boson production followed
by a $W \to \tau \nu_\tau$ decay.
To strongly suppress this large background
while keeping a sufficiently large detection efficiency for the tau leptons,
a multi-variate discrimination technique, 
called PDE-RS\footnote{Probability Density Estimation based on Range
Searching.}, was used\cite{PDE-RS}. 
Six observables were exploited to characterise the internal jet structure\footnote{
In general, the internal jet structure is well modelled by the MC
simulations\cite{jetintstructexp}.}:
the first and second moment of the radial extension of the jet\footnote{Jets were defined using the 
inclusive longitudinally invariant $k_T$ algorithm\cite{inclkt},  
$E_T > 5 \GeV$ and $-1 < \eta < 2.5$ was required.},
the first and second moment of the projection of the jet onto its axis,
the subjet multiplicity\footnote{The subjet multiplicity describes
the number of localised energy depositions within a jet that can
be resolved using a certain resolution criterion. An exact definition
can be found elsewhere\cite{jetintstructexp,jetintstructtheo}.} 
and the invariant mass of the jet four-vector, calculated from the calorimeter cells associated
to the jet.
The classification of a given event as signal or background 
was based on a discriminating variable $\mathcal{D}$. The discriminant
$\mathcal{D}$ was
calculated from the signal ($\rho_s$) and the background ($\rho_b$)
probability densities 
in the $6$-dimensional vicinity of the event to be classified:
$\mathcal{D} = \rho_s/(\rho_s+\rho_b)$.
The probability densities were sampled with MC simulations and
were calculated using a fast range-search algorithm. 
An inclusive CC DIS
Monte Carlo simulation
was used as background and a simulation
of $W^\pm$-boson production as signal process.

\begin{figure}[th]
\begin{center}
\begin{minipage}[t]{0.5\linewidth}
\psfig{figure=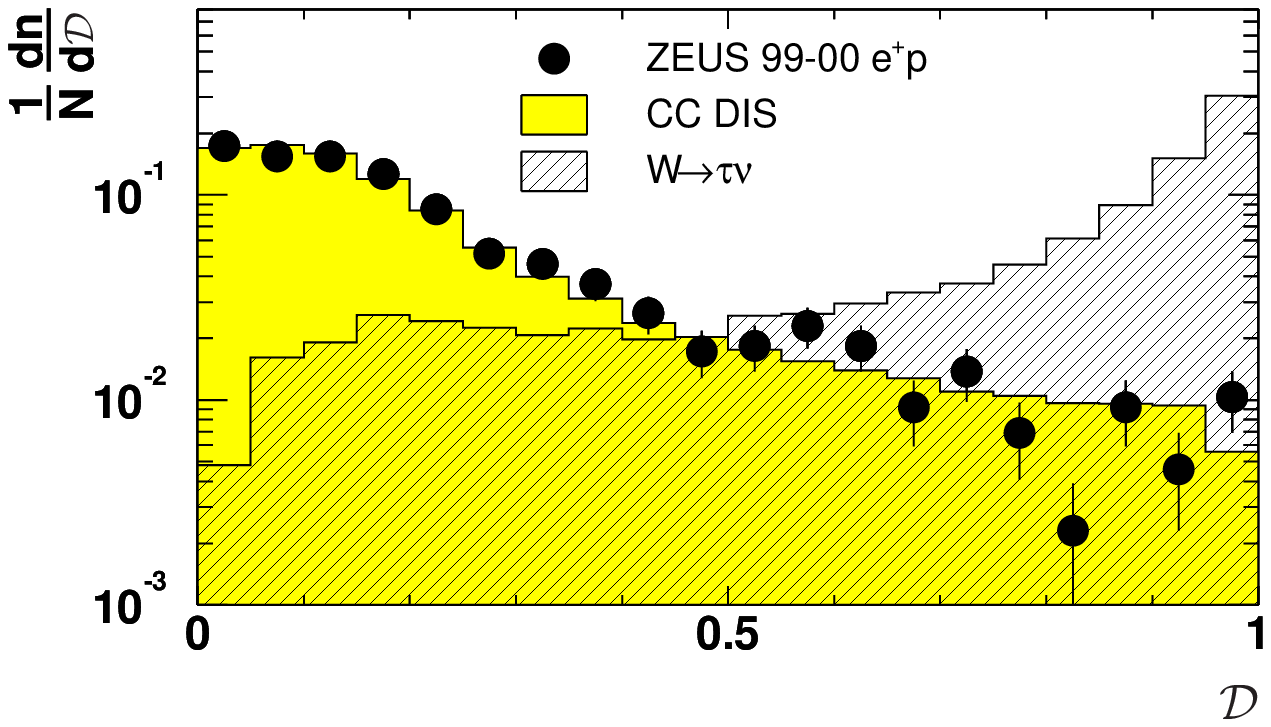, width=0.95\linewidth,clip=}
\end{minipage}\hfill
\begin{minipage}[t]{0.5\linewidth}
\psfig{figure=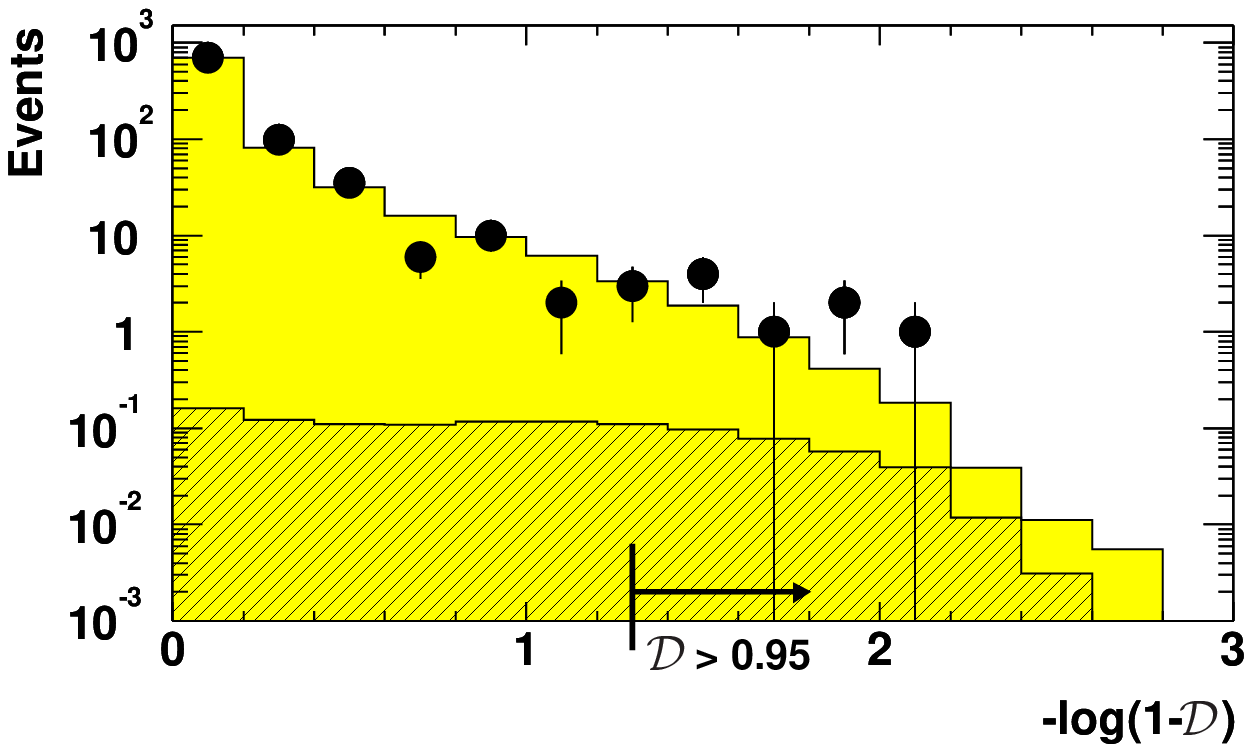, width=0.95\linewidth,clip=}
\end{minipage}
\end{center}
\begin{picture}(0,0) \put(30,108){a)} \put(222,108){b)}  \end{picture}
\vspace{-0.5cm}
\caption{
Distribution of the tau-lepton discriminant, $\mathcal{D}$, for 
an inclusive selection of CC DIS data events (dots), a 
simulation of CC DIS events (shaded histograms) and the 
simulation of the $W^\pm$-boson production
signal, where 
the tau lepton decays hadronically (hatched histograms). In each
event, only the jet with the highest $\mathcal{D}$ value enters.
The histograms are normalised (a) to the total number of events $N$ and
(b) to the luminosity of the data. 
In (b), the $-\log (1-\mathcal{D})$ distribution is displayed 
to expand the region in which the tau-lepton signal is expected.
\label{fig:zeus_tau_discri}}
\end{figure}

Fig.~\ref{fig:zeus_tau_discri}a shows the shape of the
resulting discriminant $\mathcal{D}$ for data and simulations.
The background (signal) is mostly located at low (large) $\mathcal{D}$ values.
The shape of the measured discriminant distribution is well described
by the simulation. 
When applying a cut at $\mathcal{D}>0.95$, a
signal efficiency of $\epsilon_{\rm sig}=31 \pm 0.2 \%$ and
a background rejection $R=1/\epsilon_{\rm bgd}= 179 \pm 6$ were obtained.
If in addition jets with only one track were required, the signal
efficiency was $\epsilon_{\rm sig}=22 \pm 0.2 \%$ and the
background rejection improved to 
$R= 637 \pm 41$. With this cut value for the discriminant, an optimal
separation between signal and background, 
$S=\sqrt{R}\cdot \epsilon_{\rm sig}$, was obtained.
Two alternative models were used to simulate the QCD cascade
in the CC DIS simulation. The results from above were  
independent of the model choice.
Using a NC DIS data sample, the probability to misidentify
an electron\footnote{To reject electrons, additional cuts based
on the electromagnetic energy fraction and the energy-momentum ratio
were applied.}
or a jet with one track as tau lepton was determined to be 
on the permille level\cite{damir}. The MC simulations predicted the
same misidentification probabilities.

In Fig.~\ref{fig:zeus_tau_discri}b the absolute number of 
measured and expected events
is shown. To make the region of large $\mathcal{D}$ values more visible, 
the $x$-axis was stretched according to $-\log{(1-\mathcal{D})}$.
For this inclusive CC event control 
selection, good agreement
between data and simulation was found. However, at the largest
$\mathcal{D}$ values, slightly more events were found than expected by
the SM processes.

\begin{figure}[th]
\begin{center}
\begin{minipage}[t]{0.5\linewidth}
\psfig{figure=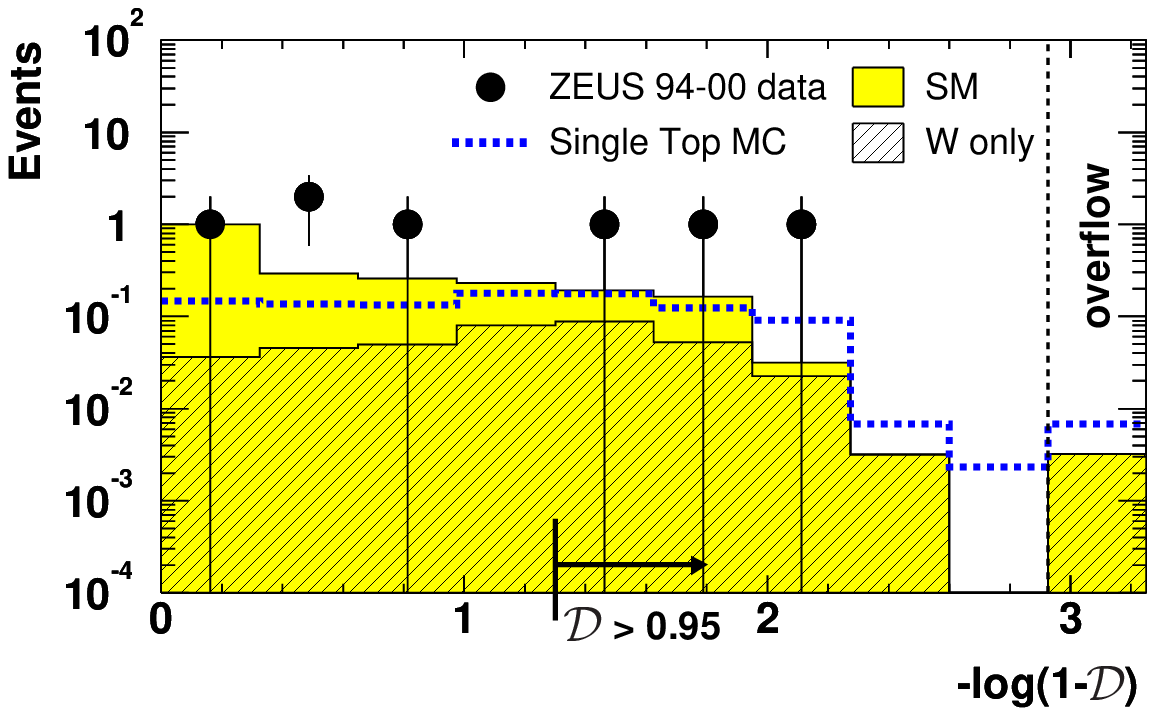, width=0.95\linewidth,clip=}
\end{minipage}\hfill
\begin{minipage}[t]{0.5\linewidth}
\psfig{figure=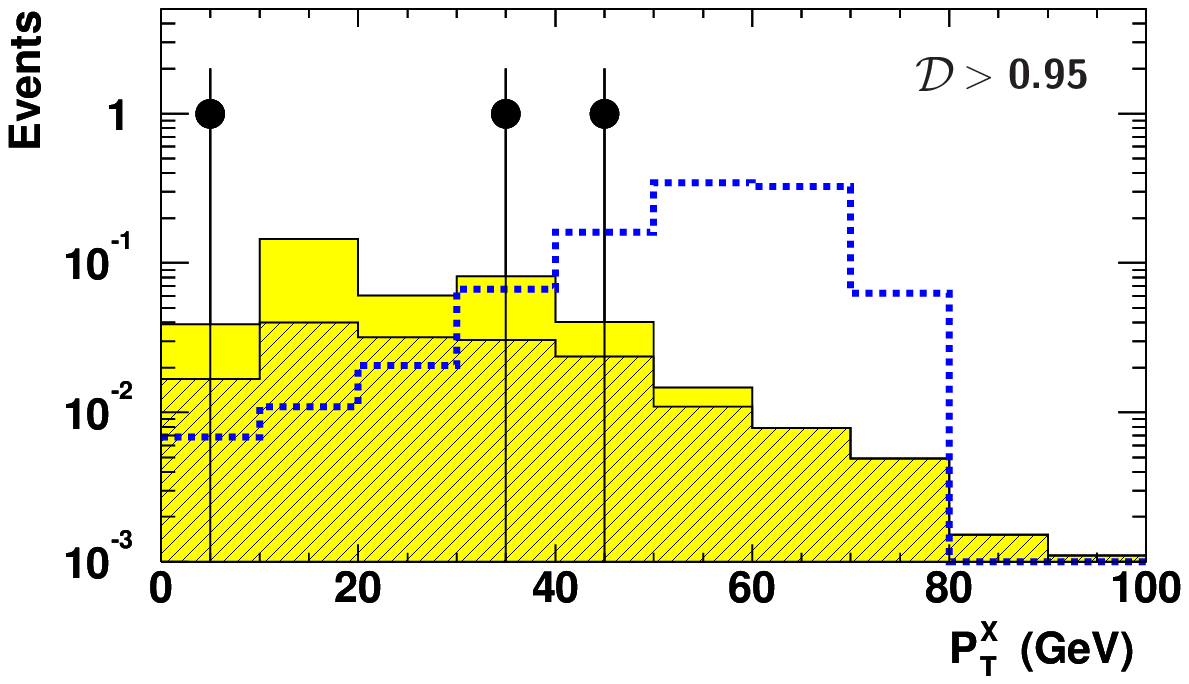, width=0.95\linewidth,clip=}
\end{minipage}
\end{center}
\begin{picture}(0,0) \put(28,108){a)} \put(220,108){b)}  \end{picture}
\vspace{-0.5cm}
\caption{
Distribution of (a) the tau-lepton discriminant, $-\log (1-\mathcal{D})$, 
before applying the cut $\mathcal{D}>0.95$ and (b)
the hadronic transverse momentum, $P_T^X$, 
after applying the cut $\mathcal{D}>0.95$.
The data (points) are compared to the SM expectations 
(shaded histogram). The
hatched histogram shows the contribution of $W^\pm$-boson production.
The dashed line shows a possible contribution
of  anomalous single top-quark production, 
normalised to an integral of one event.
\label{fig:zeus_tau-result}}
\end{figure}

\subsection{Event Selection and Results}
The event selection closely followed the ZEUS electron/muon analysis. 
The isolated track had to be associated to a tau-lepton candidate jet.
To remove background from CC DIS, the jet-isolation requirement 
was tightened to $D_{jet} > 1.8$. The final cut on $P_T^X$,
optimised for the detection of tau-leptons originating from
the decay of a heavy state, was set to $P_T^X>25 \GeVx$.
Seven events were found in the data, while only $2.2 \pm 0.5$
were expected from SM processes. Three events 
passed the cut $\mathcal{D}>0.95$ (see Fig.~\ref{fig:zeus_tau-result}a). 

\begin{figure}[ht!]
\begin{center}
\psfig{figure=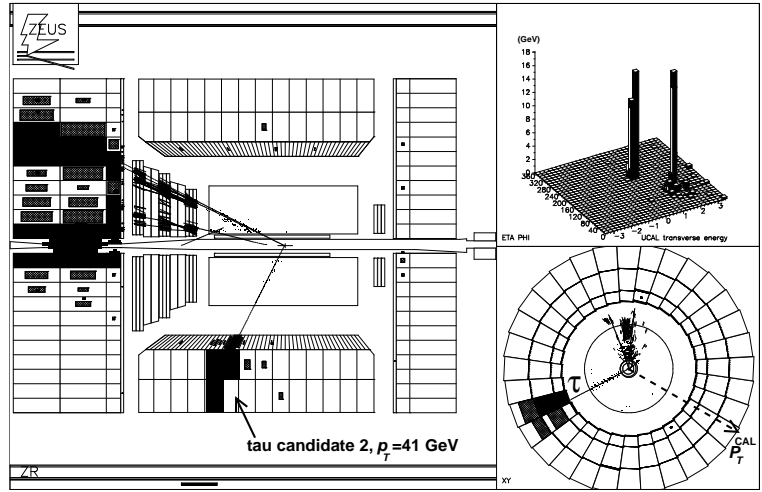,angle=270,width=10.cm}
\caption{Display of an isolated tau-lepton event with
$P_T^{\rm miss}=39 \GeV$ and $P_T^X=38 \GeV$ in the ZEUS detector.
\label{fig:zeus_tau_event}}
\end{center}
\end{figure}

For $P_T^X > 25 \GeVx$ ($P_T^X > 40 \GeVx$)
two (one) events were found in the data and
only $0.2\pm 0.05$ ($0.07 \pm 0.02$) events were expected from SM processes
(see Fig.~\ref{fig:zeus_tau-result}b). One of the data events
at large $P_T^X$ is shown in Fig.~\ref{fig:zeus_tau_event}.
The SM expectation was dominated by $W^\pm$-boson
production. The Poisson probability to observe two or more events 
when $0.2 \pm 0.05$ are expected is $1.8\%$.

\section{Search for  $W^\pm$ Bosons in the Hadronic Decay Channel}
Searches for events where the $W^\pm$ boson decays hadronically
were also performed\cite{h1,h1-singletop,ZEUS03}. However, in this
channel the SM backgrounds were so high that no firm conclusions could be
drawn. In the H1 as well as in the ZEUS analysis the measured
events were in agreement with the SM expectation dominated by QCD processes such as
$\gamma p \to {\rm jet} \; {\rm  jet}$. Anyway, since the
sensitivity in the hadronic channel was lower than in the leptonic channel,
the H1 excess in the leptonic channel is not contradicted by this result.

\section{Interpretations Beyond the Standard Model}
Several interpretations based on BSM effects have been proposed
to explain the excess of events in the electron and muon
channel measured by H1. The excess in the tau-lepton channel recently
reported by ZEUS has not yet been discussed.

\subsection{Anomalous Single Top-Quark Production}
\label{sec-stop}
\begin{wrapfigure}[21]{l}{7.5cm}
\vspace{-0.8cm}
\epsfig{figure=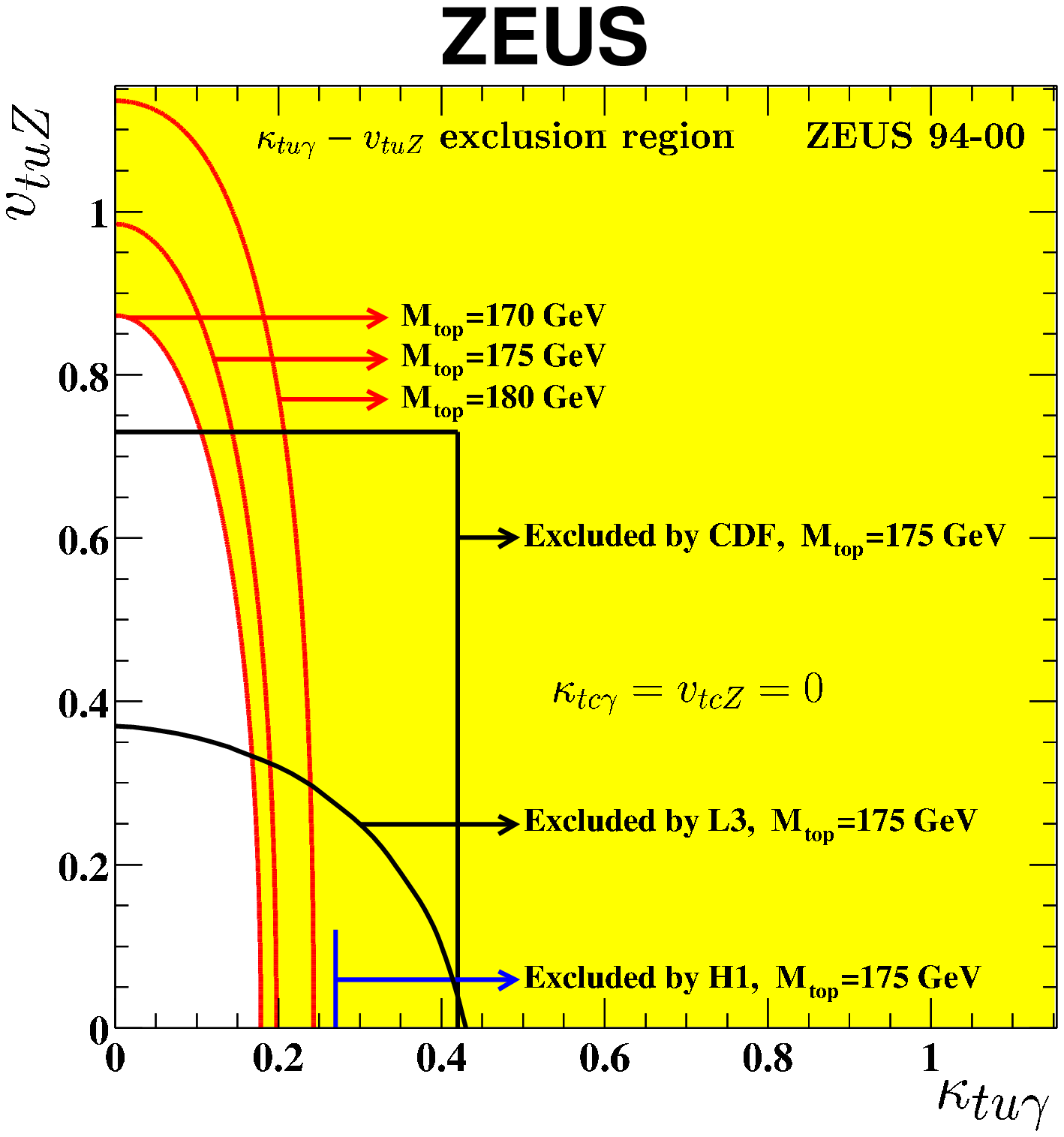,width=7.5cm}
\caption{
Exclusion regions at $95\%$ CL in the $\kg-\vz$ plane for three values
of $\mt$ assuming $\kappa_{tc\gamma}=v_{tcZ}=0$.
The CDF and L3 exclusion limits are also shown.}
\label{fig:stop2d}
\vfill
\end{wrapfigure}
The production of a single heavy particle, like a top-quark
decaying through the decay chain $t \rightarrow bW$, would
naturally lead to the observed event topology.
Within the SM, single top-quark production at the tree level can only  
proceed via a CC reaction
$ep\rightarrow \nu t\bar b {\rm X}$\cite{np:b299:21},
which, at HERA energies, has a cross section of less than
$1$\fbx\cite{pr:d56:5919}, preventing any detection
with the present integrated luminosity.
Quark flavour changing neutral current (FCNC) processes
are present only via higher order radiative corrections
and are highly suppressed due to the Glashow-Iliopoulos-Maiani (GIM)
mechanism\cite{pr:d10:1285}. 
However, many extensions of the
SM can lead to an enhancement of top production via
FCNC couplings mediated by $\kappa_{tqV}$ couplings (with $V = \gamma, Z^0$
and $q = u,c$). In order to evaluate the experimental sensitivity
to FCNC single top-quark production, a magnetic copling 
$\kappa_{tq\gamma}$ and a vector coupling $v_{tuZ^0}$ are usually 
considered\cite{pl:b457:186,pl:b224:423}.
Such couplings would induce the NC reaction $\ept$\cite{pl:b457:186,pr:d65:037501}, 
in which the incoming lepton
exchanges a $\gamma$ or $Z^0$ with an up-type quark in the 
proton\footnote{Due to the large $Z^0$-mass the coupling to the photon
is dominant at HERA. Moreover, a large value of the proton momentum 
fraction $x$ is needed to produce the heavy top quark. 
Therefore, since the proton structure function at high $x$
is dominated by valence quarks, HERA is only sensitive to
couplings involving a $u$ quark.}.


ZEUS observed no event which is compatible with anomalous
single top-quark production. H1 performed a dedicated analysis\cite{h1-singletop}
where the selection cuts were optimised for single top-quark events. 
Five events were observed, while $1.3 \pm 0.2$ were expected from SM processes.
Since no significant excess was observed, both collaborations
set limits on the coupling $\kg$ at 95$\%$ confidence level (CL), 
evaluated assuming
$\vz=0$: 
\begin{equation}
\kg < 0.27 \ (\mathrm{H1}) \;\;\;\;\;\;  \kg < 0.174 \ (\mathrm{ZEUS}). \nonumber
\end{equation}
To extract these limits also the results on hadronic $W^\pm$-boson decay 
and
NLO QCD corrections to the single top-quark 
cross section\cite{pr:d65:037501} were included.
The ZEUS collaboration also considered the effect of a non
vanishing $\vz$ coupling. In this case a LO cross section, evaluated
using the program CompHEP\cite{hep-ph:9908288},
was used, since NLO corrections are not available for the full process.
The H1 limit is less stringent because of the excess above the 
SM expectation observed in the electron and muon channels. 

Fig.~\ref{fig:stop2d} shows the 95$\%$ CL ZEUS and H1 limits in the $\kg$-$\vz$ plane,
together with the results of the L3 collaboration\cite{pl:b549:290}
obtained in $e^+e^-$ collisions at LEP\cite{pl:b494:33} 
(the results of the
other LEP collaborations are similar\footnote{
The Lagrangian used by the HERA and LEP experiments differs for a constant 
factor that leads to the following relations for the couplings 
$\kg^{\mathrm{LEP}}= \sqrt{2} \cdot \kg^{\mathrm{HERA}}$ and 
$\vz^{\mathrm{LEP}}=\sqrt{2} \cdot \vz^{\mathrm{HERA}}$.
and Tevatron, differently from the HERA experiments, had 
similar sensitivity to both $u$- and $c$-quarks.})
and of the CDF collaboration\cite{prl:80:2525} obtained
in $p \bar{p}$ collisions at Tevatron. 
The limits were reported using HERA conventions for the 
couplings and assuming vanishing couplings to $c$-quarks. 
The HERA limits are the most stringent in the region of low $\vz$.
Due to the larger sensitivity to the $Z^0$-exchange process, 
the LEP experiments obtained the most stringent limits at large $\vz$.

\subsection{{\boldmath $R$}-Parity-Violating Supersymmetry}\label{sec-susy}
Supersymmetry (SUSY) linking bosons and fermions
provides a consistent framework for the unification of the gauge
interactions (GUT). 
Assuming supersymmetric
partners for SM particles, bosonic partners for fermions and vice-versa,
it cures 
the radiative quadratic divergences to the Higgs boson mass 
stabilizing
the gap between the GUT and the electroweak scale\cite{np:b188:513}.

In the minimal SUSY extension of the SM, $R$-parity is a multiplicative
quantum number which 
is $R_p=1$ for SM particles and $R_p=-1$ for their SUSY partners\footnote{
$R$-parity is defined as
$R_p=(-1)^{3B+L+2J}$, where $B$ is the baryon number, $L$ the lepton number
and $J$ the spin of the particles.  
}.
In
$R_p$-conserving processes, SUSY 
particles are pair produced and the lightest SUSY particle (LSP)
is stable. However, 
the most general renormalizable and gauge invariant  
SUSY Lagrangian contains $R_p$-violating (\rp) terms\cite{pr:d26:287} 
that allow for single SUSY-particle production and
LSP decay into SM particles. 

Of special
interest for HERA are \rp\, Yukawa couplings that couple a squark\footnote{The SUSY
partners of the fermions are in the following called squarks, sleptons etc.,
the ones of the bosons neutralino, photino, gluino etc.. SUSY particles are denoted by a tilde.} 
to a lepton and a quark, allowing for
resonant production of squarks through $eq$ fusion\cite{pr:d40:2987}.
Such interactions are described 
in the SUSY Lagrangian by the term
$\lambda^\prime_{ijk}L^iQ^j\overline{D}^k$, where $\lambda^\prime$ is a Yukawa
coupling, $i$, $j$ and $k$ are generation indices, $L$ and $Q$ denote the left-handed
lepton and quark-doublet superfields and $\overline{D}$ denotes the
right-handed quark-singlet chiral superfield.
%

The coupling $\lambda_{1j1}^\prime$
gives rise to the reaction $e^+d\to\tilde{u}_L^j$
and the coupling $\lambda_{11k}^\prime$ to  $e^-u\to\tilde{d}_R^k$.
The $\lambda_{131}^\prime$ coupling is of special
interest since the large top mass implies a large mixing between the
left- ($\tilde t_L$) and right-handed ($\tilde t_R$) states of the
stop\cite{pl:b128:248}. As a consequence,
the two stop mass eigenstates ($\tilde t_1$ and $\tilde t_2$) are strongly non-degenerate
and $\tilde t_1$ is the best candidate to be the lightest squark.
If the mass of $\tilde t_1$ is less than $\sqrt{s}$, the lighter
stop can be resonantly produced via the $\lambda_{131}^\prime$
coupling and subsequently decay via \rp~or gauge couplings.

\begin{figure}
\vfill
\vspace{-0.25cm}
\hspace{0.25cm}
\epsfig{figure=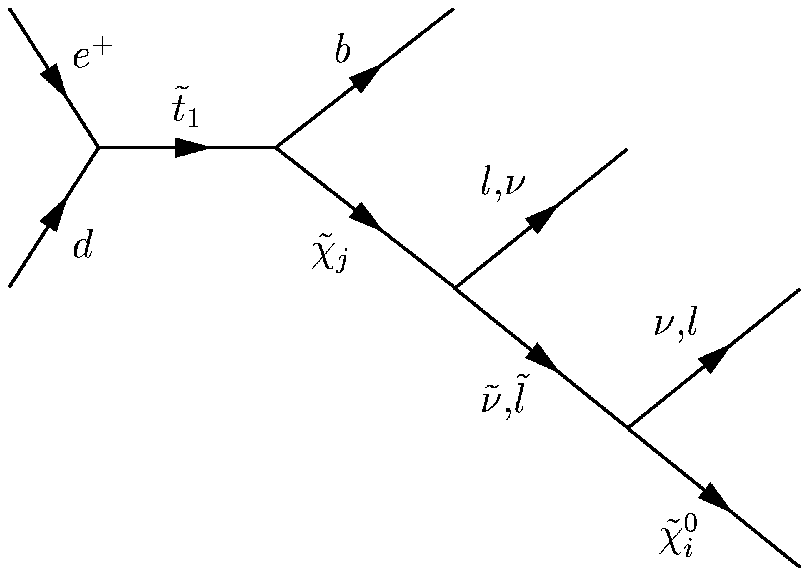,width=5cm}
\hspace{1.cm}
\epsfig{figure=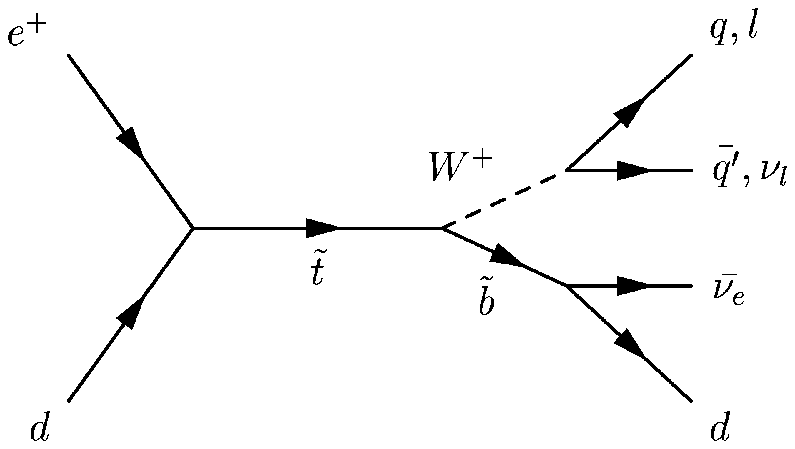,width=5cm}
\begin{picture}(0.,0.)
\put(-310,0){a)} \put(-120,0){b)} 
\end{picture}
\caption{Example Feynman diagrams for $R_p$-violating resonant squark production at HERA.}
\label{fig:feynman_rpv_susy}
\vfill
\end{figure}
Due to the large top-quark mass, scenarios exist where the two-body 
gauge decay modes involving a final-state top ($\tilde t_1 \rightarrow t
\tilde \chi^0_i$, $t \tilde g$, where $\tilde \chi^0_i$, $i=1..4$, 
and $\tilde g$ are neutralinos and gluinos, respectively)
are kinematically forbidden. In these scenarios high-$P_T$ isolated
leptons and large $P_T^{\rm miss}$ can be produced, as shown 
in Fig.~\ref{fig:feynman_rpv_susy}a, through the
decay chain $\tilde t_1 \rightarrow b \tilde \chi^+_j$,
$\tilde \chi^+_j \rightarrow \ell \tilde \nu_\ell (\nu_\ell \tilde \ell)$,
$\tilde \ell (\tilde \nu_\ell) \rightarrow \ell (\nu_\ell) \tilde 
\chi^0_i$ where $\tilde \chi^+_j$ ($j=1,2$) are charginos,
$\ell (\nu_\ell)$ and $\tilde \ell (\tilde \nu_\ell)$ indicate a charged lepton
(neutrino) and its supersymmetric partner, 
respectively~\cite{kon,pr:d59:095009}.
The charginos can be real or virtual (depending on the $\tilde t_1$ 
and $\tilde \chi^+_j$ mass difference) and the final state $\chi^0_i$ 
will decay to SM particles via an \rp~coupling.

The exact process topology crucially
depends on the SUSY parameters involved and on the relative strength between
\rp~and gauge couplings. At large values of $\tan \beta$, the ratio
of the vacuum expectation value of the two neutral Higgs fields, the mixing between
left- and right-handed states becomes relevant also in the $\tau$ sector leading to
a light mass state $\tilde{\tau}_1$ and to an enhancement of tau-lepton production with respect to
$e$ and $\mu$\cite{pr:d59:095009}. 
Considering the ZEUS excess in the isolated tau-lepton
analysis\cite{ZEUStau}, it would be interesting to study this 
large $\tan \beta$ region of the SUSY parameter space in further detail.


An anomalous $W^\pm$-boson production can
be caused by the other stop decay chain $\tilde t_1 \to \tilde b_1 W$, where $\tilde b_1$, the 
lighter sbottom quark, can subsequently decay to $\tilde b_1 \to \bar{\nu_e} d$ via 
the $\lambda_{131}$ coupling\cite{kon}.
A diagram of this process is shown in Fig.~\ref{fig:feynman_rpv_susy}b. 
The leptonic decay of the $W^\pm$ boson would then lead to a
high-$P_T$ charged lepton in the final state.
Such a process can be dominant only in a SUSY scenario where the $\tilde b_1$ is substantially 
lighter than the $\tilde t_1$. 

A dedicated search for squarks in \rp~SUSY has been performed by H1\cite{h1rp} 
including also the study of
final states with electrons or muons, multi-jets and  $P_T^{\rm miss}$. No
excess above SM expectation has been observed in this case.

\subsection{Heavy Majorana Neutrinos}
The recent observation of non-vanishing neutrino ($\nu$) masses\cite{neu_oszis} might indicate the
existence of heavy neutrino states. In most BSM scenarios, these new states are Majorana
particles ($N$) and are very massive, i.e. $M_N \approx 10^2 - 10^{18} \GeVx$, 
since they are related to the light neutrinos via the see-saw mechanism\cite{seesaw}. 

At HERA, Majorana neutrinos could be directly produced\cite{directN1,directN2,directN3} 
via the reaction $e^+ q \to N q'$.
In this case $N$ is relatively light ($M_N < \sqrt{s}$) and  
decays via $N \to W^\pm e^\mp$. The Majorana neutrino could also be exchanged in the
$t$-channel\cite{Rodejohan}, i.e. a $W^\pm$ boson is emitted from the electron beam 
and fuses with a $W^\pm$ boson from the proton beam: 
$e^+ p \to W \bar{N} W \to \bar{\nu_e} l^+_1 l^+_2 X$, where  $l^+_1$ and $l^+_2$ are
two positively\footnote{The leptons have the same charge sign as the incoming beam lepton.} 
charged leptons which can have all lepton flavours. In this case $M_N$ can be much bigger than $\sqrt{s}$.
Since one of the two leptons has a sizeable probability not to be measured in the detector, the event
topology would roughly fit to the experimental observation.
In the direct production as well as in the $t$-channel exchange 
also dilepton
events with large $P_T^{\rm miss}$ should be observable. This could allow
to distinguish this production mechanism from other possibilities. A higher event rate
is expected in $e^- p$ collisions.

The size of the coupling depends on the angle mixing the light $\nu$-state with $N$. 
However, for couplings
not excluded by low energy experiments\cite{belanger} and for reasonable values of $M_N \lesssim 1000$~GeV,
the expected cross section is tiny\footnote{Note, however, that the bounds from low energy
experiments are model dependent\cite{directN1}.}. 
For the isolated electron and muon
analyses only about $10^{-7}-10^{-8}$ events are expected for the present integrated luminosities. 
The interest in this channel arises from
the special role
of the coupling of leptons belonging to the third generation, since
limits on the mixing angles of third generation leptons are a factor of five weaker
than for the first and second generation\cite{nardi}.
Moreover,
the production with two distinct lepton flavours leading to different final states
is enhanced by a factor of about $3$ with respect to the case with the same lepton flavours.

\section{Consistency and Generic Interpretation of the Measurements}
\label{sec:inter}
The H1 experiment observed $6$ events with $P_T^X > 40$ \GeV
in the electron and muon
channel while only $1.3\pm 0.3$ events were expected from SM processes.
The Poisson probability for the SM expectation to fluctuate to this observed
number of events or more is $0.3\%$, corresponding to a significance
of $2.8$ standard deviations ($\sigma$)\footnote{The number of standard
The number of standard deviations,
$\sigma$, were obtained by relating the probability values to the case of
a one-dimensional Gaussian distribution, such that the probability for a
point to lie outside the quoted number of $\sigma$ to one side of the mean
is given.}.
The ZEUS results are in good agreement with the SM. 
However, ZEUS
observed $2$ events  $P_T^X > 25$ \GeV
in the tau-lepton channel, where only $0.2 \pm 0.05$ events
were expected, corresponding to an excess
of approximately $2\sigma$. 
So far, H1 has not released results in the tau-lepton channel. 

Monte Carlo simulations 
were
performed to evaluate whether the event yields, observed by the two
experiments in the different channels, are compatible within
SM background expectations or whether hypothetical BSM processes
can lead to a more consistent interpretation\cite{dominik}.
In the following, three possible BSM scenarios are considered.
Each scenario was characterized by an additional cross section, $\sigma_{\rm BSM}$,
leading to an extra-production of isolated leptons.
A large number of randomly simulated experiments have been generated for 
each experimental result. Each simulated experiment was characterized by a
number of observed events, $N^{\rm MC}_{obs}$, generated from a 
Poissonian distribution with expectation value given by the
sum of the SM expectation and the BSM contribution.
The compatibility between the experimental results and the BSM
scenario considered was studied varying $\sigma_{\rm BSM}$ and was measured by
the variable $P_{\rm obs}(\sigma_{\rm BSM})$, defined as the fraction of randomly simulated
experiments, which had a lower Poisson probability than the actual
observation.
 The uncertainty in the SM expectation was
taken into account in evaluating the Poisson probabilities.
 
In the following, anomalous production of $W^\pm$ bosons, single top-quark
production and anomalous production of tau leptons are considered 
as generic models for signal processes resulting in final states with isolated
leptons (see Fig.~\ref{fig:feynman_isolept_bsm}).

\subsection{Anomalous $W^\pm$-Boson Production}
An anomalous $W^\pm$-production signal process with
cross section $\sigma_\mathrm{W,BSM}$ would, via a subsequent leptonic decay of
the $W^\pm$ boson, lead to events with isolated electrons, muons or tau leptons,
as sketched in Fig.~\ref{fig:feynman_isolept_bsm}a. The detection efficiency for 
such a process was estimated from the efficiency 
for detecting SM $W^\pm$-boson production. 
%
Figure~\ref{fig:h1zeusprob}a shows the values of $P_\mathrm{obs}$ for the
individual channels as a function of the hypothetical cross section 
$\sigma_\mathrm{W,BSM}$. The maxima of the curves correspond to the most probable
cross sections. For the ZEUS electron/muon channel, where no events were observed,
the highest probability was obtained for no 
additional $W^\pm$-boson production cross section ($\sigma_\mathrm{W,BSM}=0$),
while the H1 electron/muon analysis reached the largest value of $P_\mathrm{obs}$ 
for a cross section of $\sigma_\mathrm{W,BSM} \approx 6 \pbx$. 
Due to the reduced 
efficiency
in the ZEUS tau-lepton channel, it is most compatible with an even higher cross section of
$\sigma_\mathrm{W,BSM} \approx 23 \pbx$. The probabilities of all 
possible combinations of the three search channels are shown in Fig.~\ref{fig:h1zeusprob}b.
The values for the most probable cross sections lie in between the ones for the individual
channels. For the combination of all search channels, a value of 
$\sigma_\mathrm{W,BSM}=2.8 \pbx$ has the highest probability of $P_\mathrm{obs}=0.6\%$.
This still very low joint probability indicates that
the set of observations hardly can be an outcome of the considered
scenario.
The probability assuming no additional cross section had a value
of approximately $P_\mathrm{obs}=0.03\%$, corresponding to $3.4\sigma$ 
significance for a deviation from the SM. 

\subsection{Anomalous Single Top-Quark Production}
A more specific model assumption on the anomalous $W^\pm$-boson production mechanism was
made by comparing the observed event yields to the prediction from single top-quark
production with cross section $\sigma_\mathrm{sing.top}$. In case of H1, the 
results from the dedicated search for single top-quark
production\cite{h1-singletop} were considered, where five events were found 
while $1.3\pm 0.2$ events were expected from SM background. 
The efficiency for single top-quark production in the ZEUS tau-lepton channel
was approximately 25 times lower than in both the H1 and ZEUS combined electron/muon 
channels.
Figure~\ref{fig:h1zeusprob}c and Fig.~\ref{fig:h1zeusprob}d show the values of $P_\mathrm{obs}$ for 
the individual search channels and for all combinations as a function of 
$\sigma_\mathrm{sing.top}$. As expected from
the low efficiency, the most probable cross section for the ZEUS tau-lepton channel
reached a very large value of $\sigma_\mathrm{sing.top}=5 \pbx$.
The combination
of the ZEUS and H1 electron/muon channels alone, on the other hand, obtained a
higher value for the maximum of $P_\mathrm{obs}=6\%$, since the H1 excess 
was lower in the dedicated single-top quark search compared to the generic search for 
isolated leptons.
For the combination of all search channels, a value of 
$\sigma_\mathrm{W,BSM}=0.2 \pbx$ had the highest probability of $P_\mathrm{obs}=0.6\%$.
The probability assuming no additional cross section had a value
of $0.1\%$, corresponding to approximately $3\sigma$ significance.

\subsection{Anomalous Tau-Lepton Production}
\begin{figure}[p!]
\begin{center}
\psfig{figure=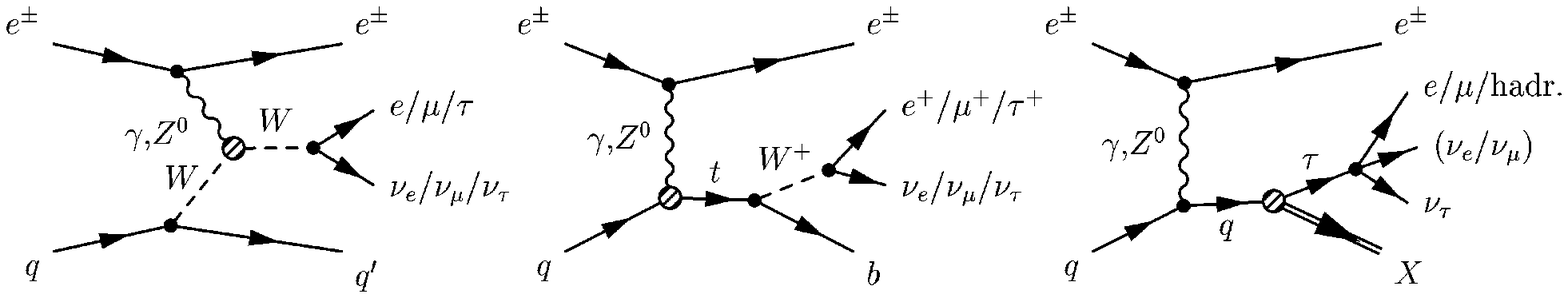,width=\linewidth}
\begin{picture}(0.,0.)
\put(-180,45){a)} \put(-60,45){b)}  \put(60,45){c)}
\end{picture}
\vspace*{-8pt}
\caption{Diagrams for generic BSM processes which could lead to final states with isolated 
leptons: Anomalous $W^\pm$-boson production (a), single top-quark production (b) and
anomalous tau-lepton  production (c).}
\label{fig:feynman_isolept_bsm}
\vspace*{15pt}
\psfig{figure=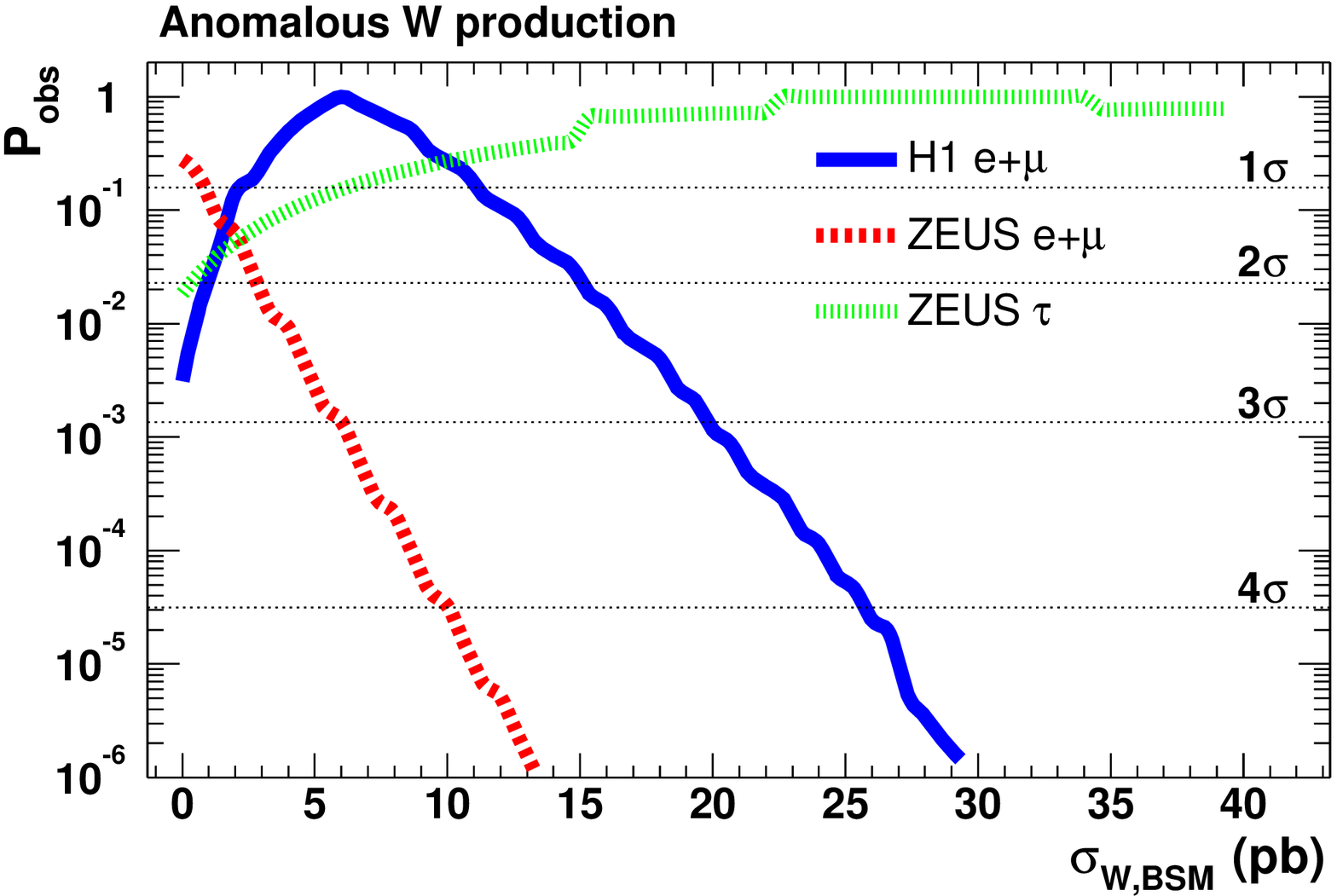,width=0.49\linewidth}
\hspace{-0.2cm}
\psfig{figure=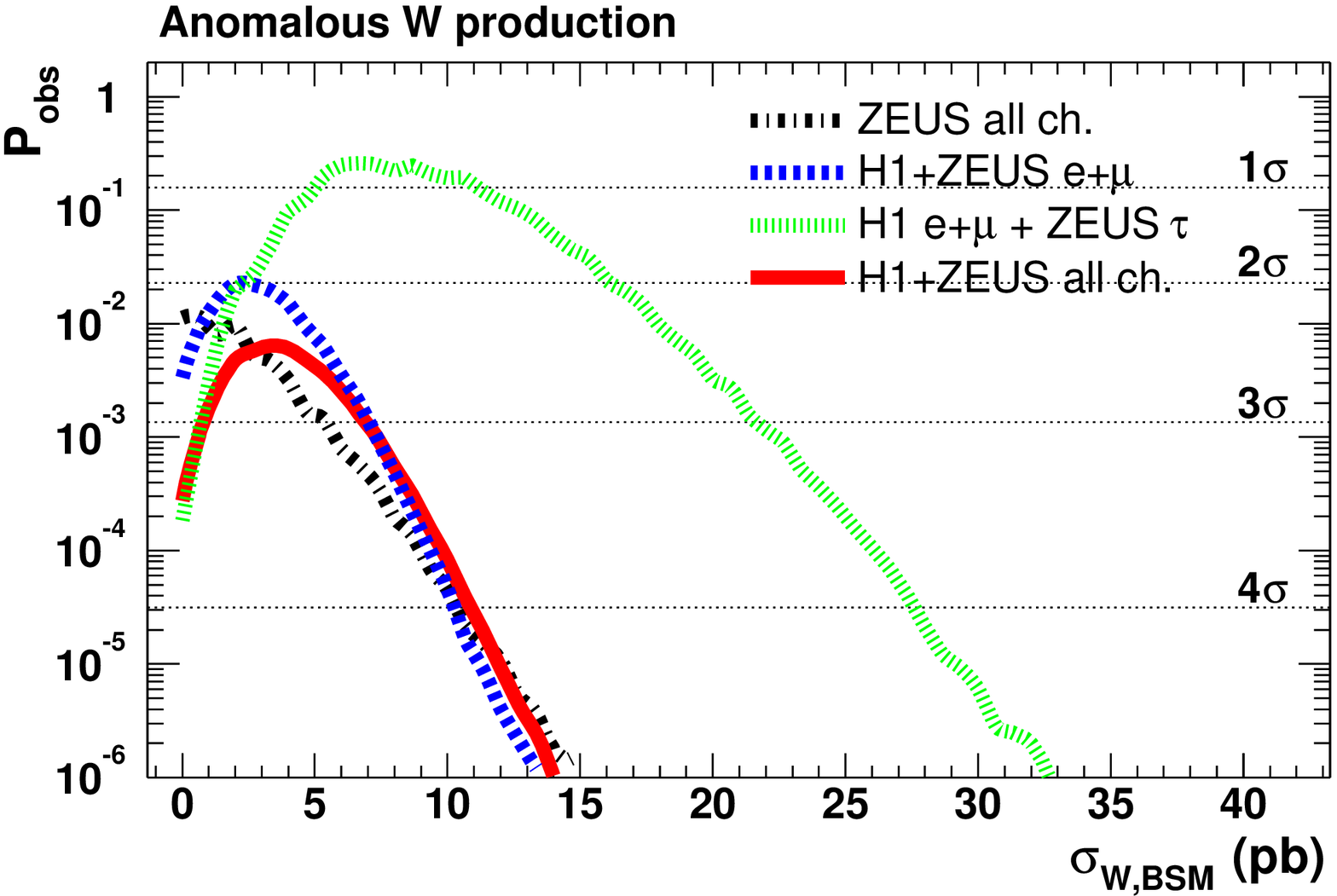,width=0.49\linewidth}
\psfig{figure=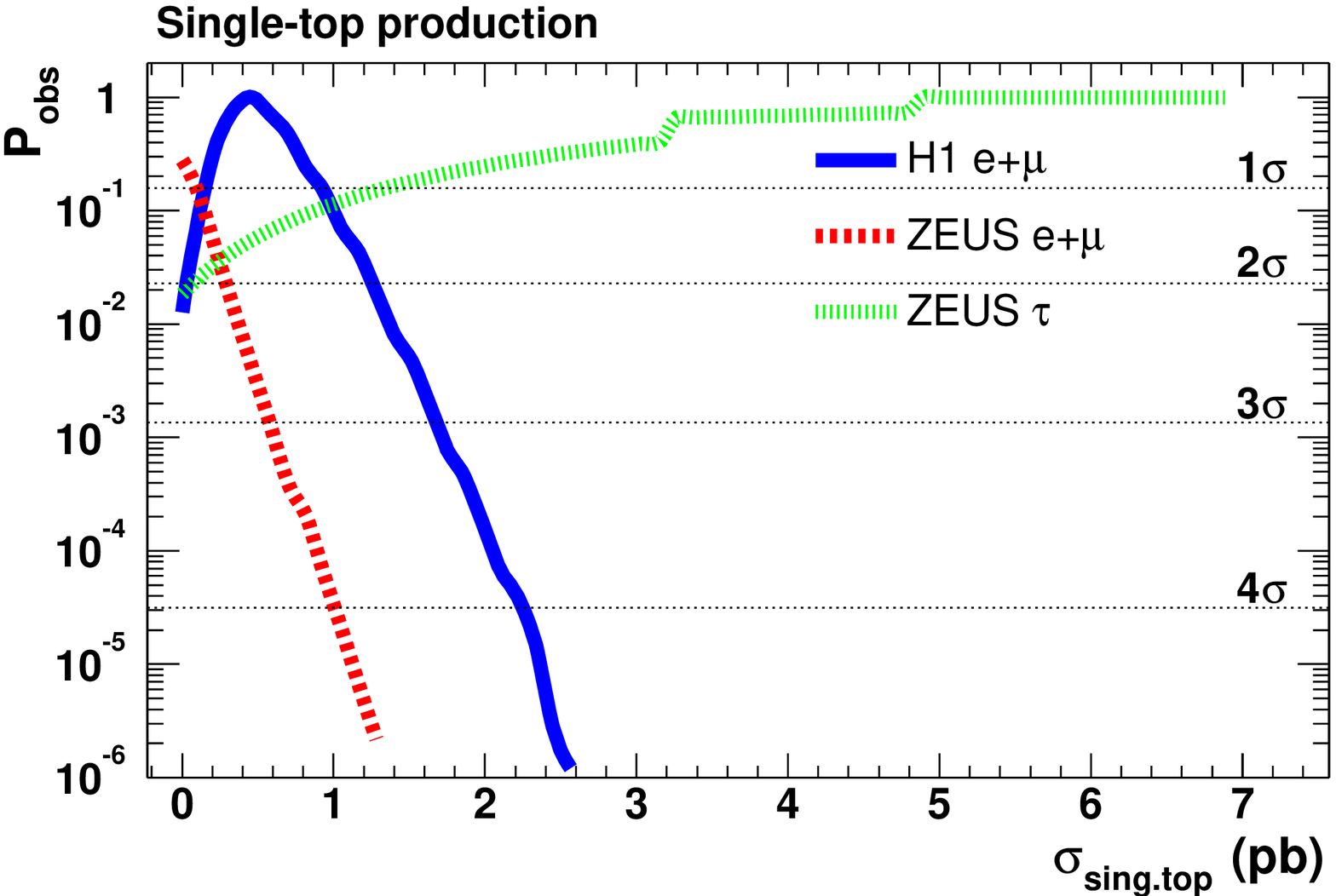,width=0.49\linewidth}
\hspace{-0.2cm}
\psfig{figure=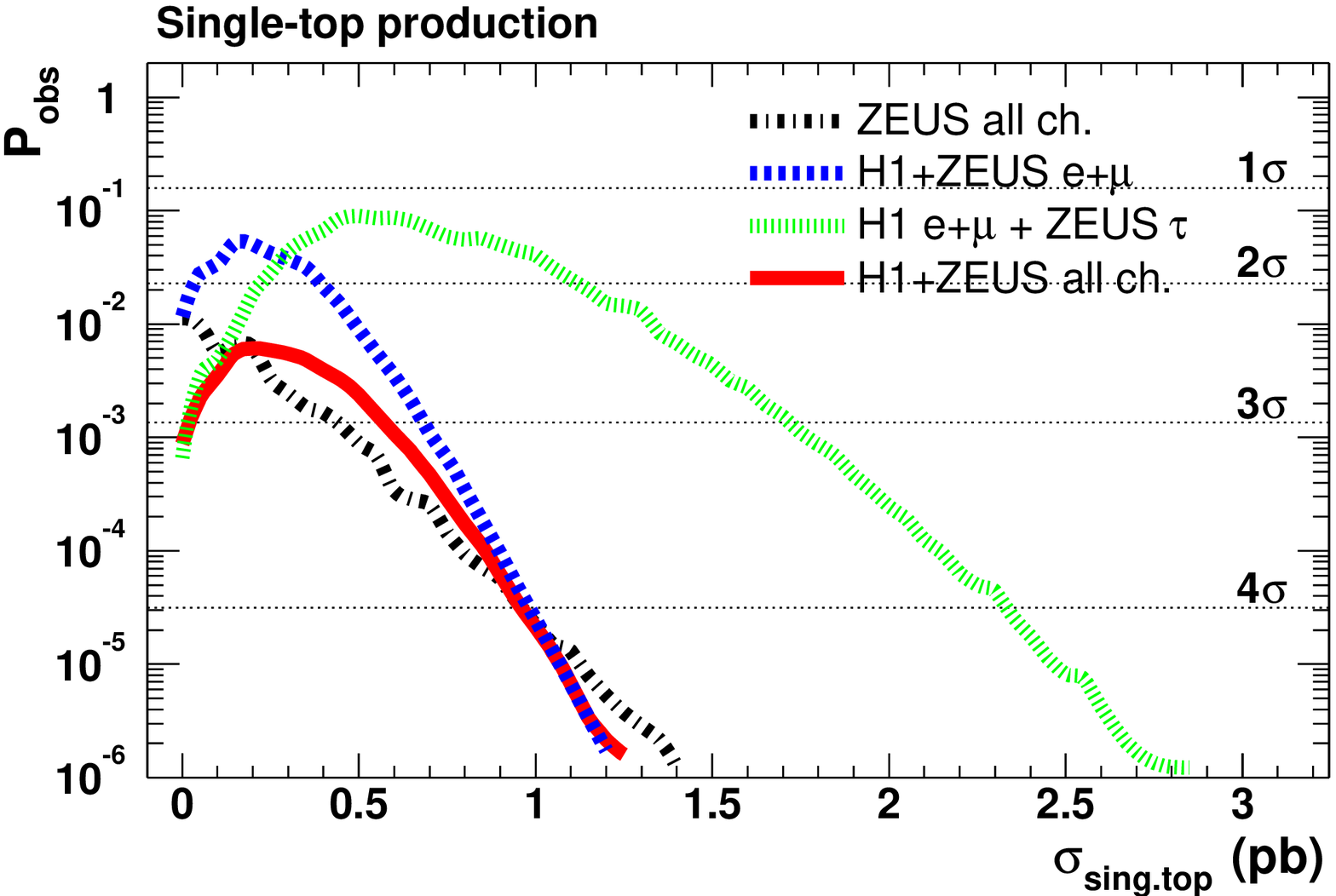,width=0.49\linewidth}
\psfig{figure=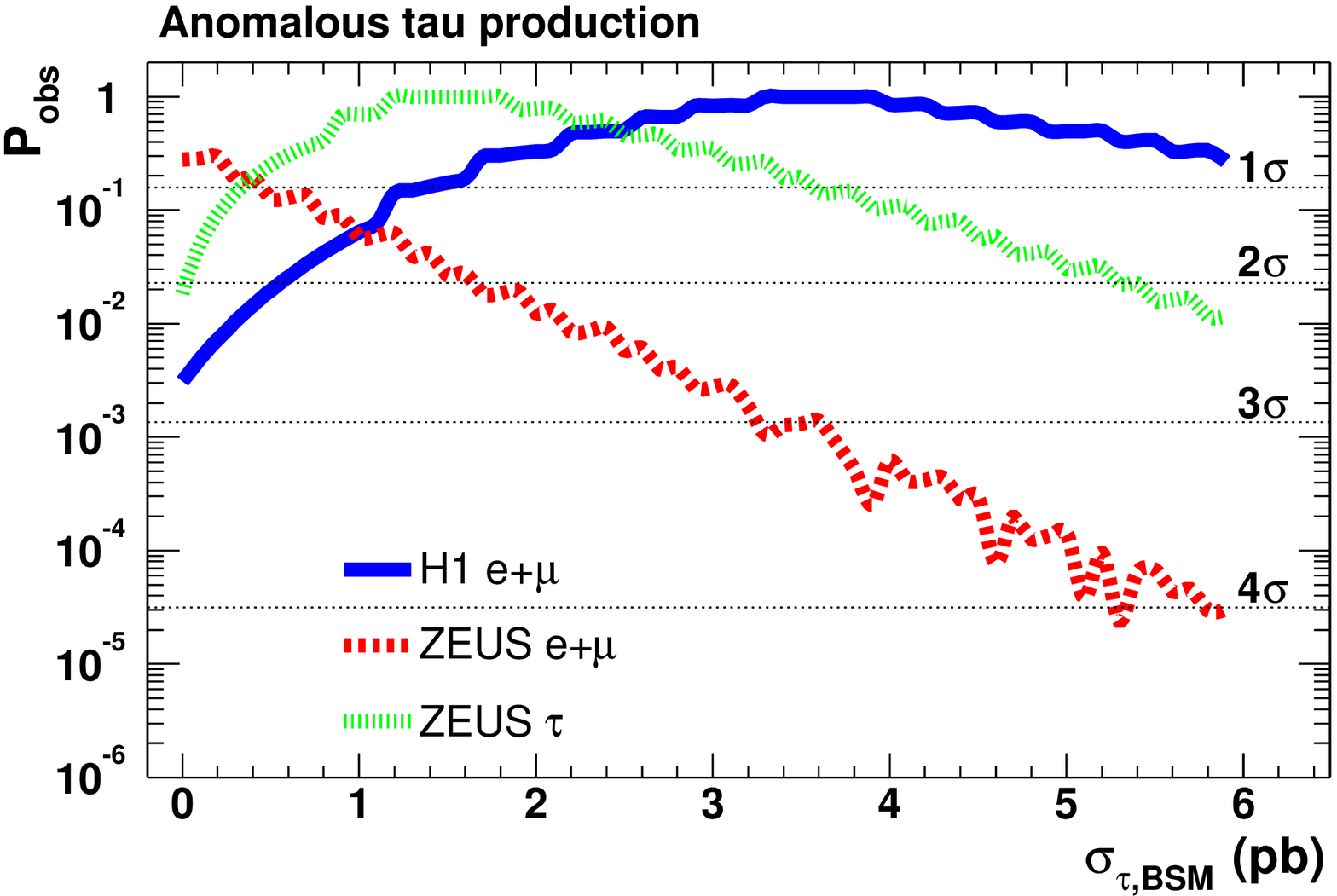,width=0.49\linewidth}
\hspace{-0.2cm}
\psfig{figure=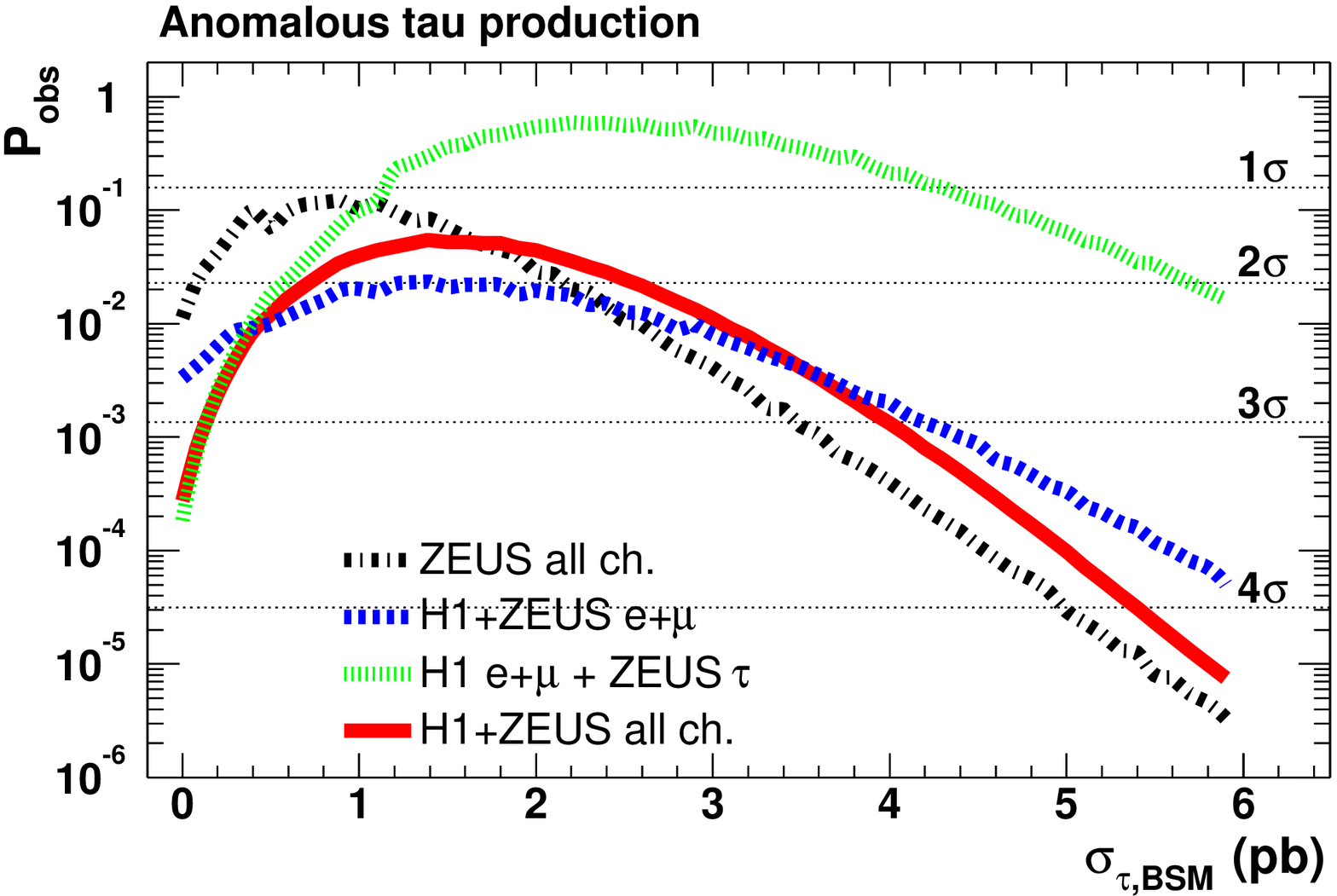,width=0.49\linewidth}
\begin{picture}(0.,0.)
\put(-347,351){a)} \put(-170,351){b)}
\put(-347,234){c)} \put(-170,234){d)}
\put(-347,111){e)} \put(-170,111){f)}
\end{picture}
\caption{Probabilities of the observed events yields, $P_\mathrm{obs}$,
for all combinations of the H1 and ZEUS search channels, as function
of additional hypothetical cross sections for three generic BSM processes:
anomalous production of $W^\pm$-bosons, single top quarks and tau-leptons.
The fluctuations in the curves are caused by discontinousities of the Poisson
probabilities, which were calculated for discrete event numbers.
}
\label{fig:h1zeusprob}
\end{center}
\end{figure}

Anomalous direct production of tau-leptons in a process beyond the SM, e.g. as
considered in section~\ref{sec-susy}, could be the cause of both, the
two tau-lepton events observed in the ZEUS tau-lepton search and the excess of electron
and muon events in the H1 analysis. 
In this scenario, the tau-leptons do not originate from $W^\pm$-boson decays
and hence the expected signal in the electron/muon channels consists only of
leptonic tau-lepton decays. Assuming that the selection efficiency
for those electrons and muons would be identical to the one for electrons and
muons from SM single $W^\pm$-boson production, the efficiency 
(inlcuding the branching ratio) to identify tau-leptons in the hadronic channel
was approximately the same as in the combined electron/muon 
channels.  
Figure~\ref{fig:h1zeusprob}e and Fig.~\ref{fig:h1zeusprob}f show the values of $P_\mathrm{obs}$ for 
the individual search channels and for all combinations as function of an
anomalous tau-lepton production cross section $\sigma_\mathrm{\tau,BSM}$. 
The probabilities assuming no additional cross section were identical to the ones
from the first scenario (anomalous $W$-boson production), since the same observations and
background expectations were considered. The combination of all search channels
had a value of $0.03\%$.
The most probable cross section value of $\sigma_\mathrm{\tau,BSM}=1.5 \pbx$ 
corresponds to two expected signal
events in the ZEUS tau-lepton channel as well as in the H1 and ZEUS combined electron/muon
channels. The probability of $5\%$ is higher than for the other cases.
All results would be compatible on a level of better than two standard deviations.
Therefore an anomalous tau-lepton production process
could provide a more consistent interpretation of all measurements
than the other scenarios considered.
\section{Outlook}
\label{sec:outlook}
\subsection{Isolated Lepton Events}
\begin{wrapfigure}[21]{l}{7.5cm}
\vspace{-0.6cm}
\psfig{figure=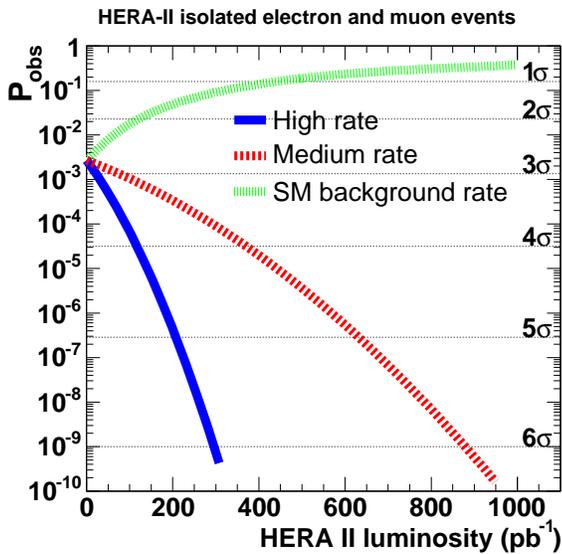,width=\linewidth}
\caption{Poisson probabilities of expected future event yields in the combined H1 and
ZEUS isolated electron and muon searches 
of the future HERA-II data taking for three scenarios, as explained in the text.}
\label{fig:isolept_future}
\end{wrapfigure}

The high luminosity expected for the HERA-II data taking period
can help to clarify whether the excess of measured events in the
electron, muon and tau-lepton channels over the SM expectation
were due to a statistical fluctuation or to a new BSM interaction.
Figure~\ref{fig:isolept_future} compares different scenarios for
the expected event
yields in future H1 and ZEUS searches
for isolated electron and muon events. It was assumed that the background 
rates and their uncertainties, 
scaled to the additional luminosity,
and the signal efficiencies will be the same as in 
the 1994-2000 analyses. 
%
Different assumptions on the future event rates $R_\mathrm{fut}$ were 
made. The probability $P_\mathrm{obs}$ was defined as the fraction 
of simulated SM-background-rate experiments, which had a lower Poisson 
probability than the actual simulated observation with rate $R_\mathrm{fut}$.
The probabilities $P_\mathrm{obs}$ are 
displayed as a function of the additional HERA-II luminosity. 
In the first scenario it was assumed that both experiments will observe events
at a high rate, corresponding to the H1 rate in the 1994-2000
data taking periods. After an additional luminosity of
$200 \pbinv$, a deviation from the SM expectation with a significance
of approximately $5\sigma$ will be observed in this
case.
The second scenario assumed that both experiments will observe
events at the average of the previous H1 and ZEUS rates.
In this case, the significance for a deviation from the SM will
be about $3 \sigma$ after an additional luminosity of $200 \pbinv$,
while after $600 \pbinv$ the significance will increase to approximately
$5 \sigma$.
In the third scenario, only the SM background rate will be observed
in both experiments. In this case, the combined H1 and ZEUS results
will reach agreement with the SM within $1-2\sigma$
after approximately $150 \pbinv$.
The sensitivity can be further increased, if the tau-lepton 
channel is also considered and if the improvements in both the H1 and
ZEUS detectors are exploited.

\newpage
\subsection{Single Top-Quark Production}
\begin{wrapfigure}[17]{l}{6cm}
\vspace{-0.8cm}
\epsfig{figure=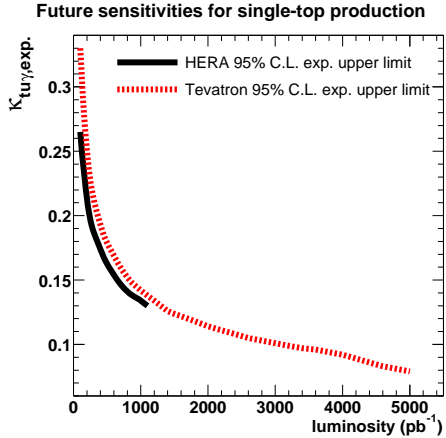,width=6cm}
\vspace{-0.2cm}
\caption{Expected future sensitivities of the HERA and Tevatron searches for
single top-quark production.}
\label{fig:singletopfuture}
\vfill
\end{wrapfigure}
The future sensitivity of the HERA experiments for anomalous single top-quark production 
was estimated from a simulation of the expected event yields in the ZEUS 
single-top search, 
assuming 
that only background processes will be measured.
The anomalous vector
coupling to the $Z^0$-boson, $\vz$, was assumed to be vanishing and LO calculations 
for the single top-quark cross section were used in the simulations.
Figure \ref{fig:singletopfuture} shows
the resulting expected 95\%~C.L. upper limit on the photon coupling, $\kg$, as a function of the
additional HERA-II luminosity. For comparison, also the expected sensitivity of the
Tevatron experiments is shown, estimated from the present
CDF results\cite{prl:80:2525} by scaling the background expectations to the additional 
luminosity expected from their Run-II.  
For an additional luminosity 
of $500 \pbinv$, the HERA experiments will be sensitive down to a coupling strength 
of $\kg \lesssim 0.16$.
The Tevatron experiments will have a similar sensitivity.

\section{Conclusion}
\label{sec:conclusion}
Both, the H1 and ZEUS collaborations observed more events with isolated
leptons, large $P_T^{\rm miss}$ and large $P_T^X$ than expected from the
SM. While H1 observed an excess of approximately three standard deviations  
in the combined electron and muon channels, ZEUS saw agreement with the SM 
in these channels. In the tau-lepton channel, however, which has only been investigated
by the ZEUS collaboration so far, two events were observed.
This corresponds to
an excess of approximately two standard deviations. 
The probability that these measurements are simuntaneously realised in random
experiments is $0.03\%$, i.e. a $3.4\sigma$ deviation from the SM expectation
was observed. The deviation is large, however, the lack of agreement between the 
two experiments in the individual search channels complicates a consistent
interpretation of the results.

Several models beyond the SM could explain
an excess of isolated lepton events as originating from the decays of heavy states. 
For certain processes, like e.g. photon mediated anomalous single top-quark production,
the HERA experiments have a higher sensitivity than any other collider.
%
%
The addition of an anomalous production of single top-quarks or $W^\pm$-bosons
does not significantly improve the consistency of the measurements.
However, an anomalous production of tau leptons with a cross section of about 1\pb
could, through the leptonic and hadronic decays, consistently explain all the 
measurements.
\rp~SUSY models provide scenarios allowing for such a BSM tau-lepton production.

The HERA-II data taking period will help to clarify the
origin of the observed excess of isolated lepton events, if an additional 
luminosity of 150-600\pbinv~can be accumulated. The sensitivity for anomalous single top-quark
production will in this case improve to values of $\kg \lesssim 0.16$.

\section*{Acknowledgements}
We would like to thank W. Buchm\"uller, C. Diaconu, E. Gallo, M. Kuze and 
F. Schrempp for critical reading of the manuscript and helpful comments.

\end{document}